\documentclass[]{aa} 
\usepackage{graphicx}
\usepackage{natbib}

\newcommand{\Teff}{\hbox{$T\sb{\rm eff}$}}          
\newcommand{\logg}{\hbox{$\log g$}}
\newcommand{\Msun}{\hbox{M$\sb{\odot}$}}

\begin{document}

   \title{New DQ white dwarfs in the Sloan Digital Sky Survey DR4:
   confirmation of two sequences}

   \author{D. Koester
	   \and 
           S. Knist
           }
   \institute{   Institut f\"ur Theoretische Physik und Astrophysik, 
             University of Kiel, {D-24098 Kiel}, Germany
}

   \offprints{D. Koester\\ \email{koester@astrophysik.uni-kiel.de}}

   \date{}

\authorrunning{D. Koester \& S.Knist}

\titlerunning{New DQ white dwarfs in the SDSS}

\abstract{Using photometric selection criteria, and in a final step
     visual inspection of spectra, we find 65 DQ white dwarfs in the
     Data Release 4 of the Sloan Digital Sky Survey. 37 of these are
     new identifications, three more are reclassified DQ from DC, the
     other 25 have been described as DQ before in the literature. We
     derive effective temperatures and carbon abundances for 60 of
     these stars. The results confirm the main conclusions of
     \cite{Dufour.Bergeron.ea05}. The majority of stars defines a
     clear sequence in the C abundance vs. \Teff\ diagram, with high
     abundances found at high \Teff\ and vice versa. We also confirm
     with high significance a second sequence with an abundance about
     1 dex higher at the same \Teff, and discuss the nature and
     possible origin of the high-C sequence.   }

\maketitle

\section{Introduction}
White dwarfs of spectral type DQ are defined as showing atomic or
molecular features of carbon in a helium atmosphere. Contrary to other
metals (notably Ca) found in cool white dwarfs, which are thought to
be provided through an accretion process from outside the star, the
most widely accepted explanation for the DQs is dredge-up from the
underlying carbon/oxygen core through the expanding He convection zone
\citep{Koester.Weidemann.ea82, Pelletier.Fontaine.ea86}. A detailed
summary of the basic properties of this class and earlier references
to spectroscopic analyses has been given recently in a comprehensive
paper by \citet[][= DB05 henceforth]{Dufour.Bergeron.ea05} and need
not be repeated here.

In addition to the general understanding of the nature, evolutionary
relations and origin of the peculiar spectral types of white dwarfs,
the DQs are of special interest, because they provide information
about the deeper layers of the stars. The transition between the outer
helium layer and the carbon/oxygen core -- the result of the helium
burning in the progenitor -- is not abrupt, but rather gradual as
determined by the diffusion equilibrium
\citep{Koester.Weidemann.ea82}. We can predict the structure of this
transition quantitatively \citep{Pelletier.Fontaine.ea86}, as well as
the depth of the helium convection zone at a given effective
temperature and mass of the star. The observed carbon abundance in the
atmosphere is then directly related to the thickness of the helium
layer remaining from the previous nuclear evolution, which is a very
important parameter for the understanding of stellar evolution.  The
first estimate of this thickness by \cite{Pelletier.Fontaine.ea86} --
$\log M_{He}/M \approx -3.75$ -- was much thinner than predicted by
evolutionary calculations of the AGB phase. However, the recent
abundance determinations by DB05, together with the new models of the
outer layers by \cite{Fontaine.Brassard05} are in much better
agreement with theoretical expectations. Nevertheless, there are still
many problems understanding the detailed connection of the DQ stars to
possible helium-rich progenitors \citep{Althaus.Serenelli.ea05,
  Scoccola.Althaus.ea06}. 

When searching for correlations of DQ parameters, e.g. between
atmospheric carbon abundance and effective temperature, it is very
helpful to have available a homogeneous sample of observations. It has
been found in previous studies that abundances determined from atomic
lines in the UV or molecular bands in the optical may sometimes differ
significantly \citep[e.g.][]{Provencal.Shipman.ea02}, possibly
blurring any correlation if only one or the other observation is
available for different objects \citep{Weidemann.Koester95}. Such a
homogenous sample is currently provided by the Sloan Digital Sky
Survey, and DB05 provides the first quantitative analysis of 40 DQs
from the Data Release 1 (DR1). In this paper we report similar results
for 40 new DQs (and 20 previously known) extracted from DR4
\citep{Adelman-McCarthy.Agueros.ea06}.

\section{Selection of DQ candidates}
DR4 of the SDSS contains spectra for approximately 240000 objects
classified as ``point source'' or ``unknown''.  In order to narrow
down this huge database to tractable numbers we have applied three
consecutive selection steps, the first two based on the SDSS
photometry, the last on the spectra.

\subsection{DQ regions in SDSS color space and theoretical models}
Theoretical SDSS colors can be calculated from synthetic spectra by
convolving them with the 5 $ugriz$ SDSS band passes. Our model grid is
similar to that described in \cite{Carollo.Koester.ea03}, but extended
to cover effective temperatures from 13000~K to 4400~K. Logarithmic C
abundance by number relative to He ( = [C/He])range from $-8$ to $-4$
in steps of 0.5. The standard grid used here has a surface gravity of
\logg\ = 8.00, but additional grids were calculated with 7.5 and 8.5
for comparison. The models are fully blanketed, fully including the line and
molecular band absorption in the calculation of the
atmospheric structure. Non-ideal effects in the form of a lowering of
the dissociation energy of the C$_2$ molecule is included in a very
approximate way.

The transmission curves for the five filters (for airmass 1.3 as
recommended) were obtained from the SDSS web sites; the zero-points for
the magnitudes were set to correspond to the AB system of magnitudes
\citep{Fukugita.Ichikawa.ea96} except for small changes to $u,i,z$
(-0.04, 0.01, 0.02) similar to those discussed in
Eisenstein et al. (2006).

\begin{figure}
\includegraphics[width=8.8cm]{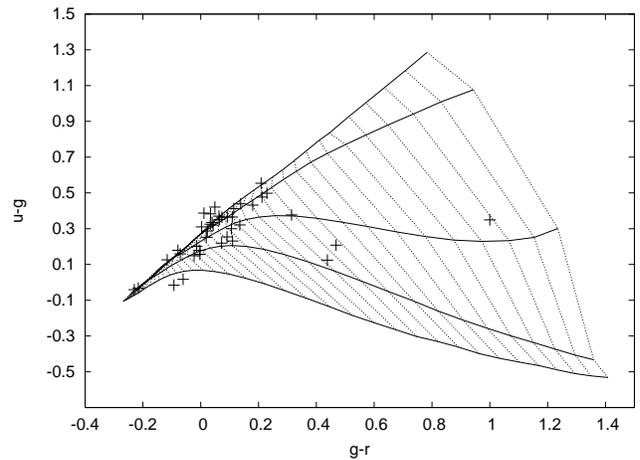}
\caption{Theoretical SDSS color grid for DQ white dwarfs. Thick
  continuous lines are lines of constant carbon abundance from [C/He]
  = -8.0 (top) to $-4$ (bottom), step 1.0. Thin dashed: lines of
  constant \Teff\ from 4400~K (right) to 13000~K (left) in steps of
  200~K. Crosses show the observed DQ stars from
  \citet{Harris.Liebert.ea03} and \cite {Kleinman.Harris.ea04}
  \label{figugr}}
\end{figure}

The dependence on \Teff\ and [C/He] is strongest in the $u-g$
vs. $g-r$ two-color diagram and much less in the remaining independent
indices used, $r-i$ and $i-z$. Figure~\ref{figugr} shows this diagram
with the theoretical grid, and 40 observed DQs from
\cite{Harris.Liebert.ea03} and \cite{Kleinman.Harris.ea04}, which
indeed fall in or very near the theoretical grid. Note that we have
not applied any correction for galactic extinction (see DB05).

\begin{table*}[ht!]
\caption{40 new DQ white dwarfs. The first
    columns give the SDSS name and internal designation. Next is
    effective temperature determined from a fit to the photometry. The
    carbon abundance [C/He] is determined from a model fit to the
    spectrum, keeping the temperature fixed. The final column gives a
    classification and remarks. WD\,1105+412 and WD\,0913+103 are reclassified from
    a DC classification in \cite{McCook.Sion99}. GD\,311 is classified
    DC in \cite{Kawka.Vennes06}, but we believe very weak Swan bands
    to be present.}
    \begin{center} \scriptsize
      \begin{tabular}{l|rrr|ccc|c}
        \hline
	\noalign{\smallskip}
        SDSS Name&MJD&Plate&Fid&$ \Teff [K]$& [C/He] &$ \sigma$
        [C/He]&Spectral Type/Ref\\ 
	\noalign{\smallskip}
        \hline
        SDSS\,J074204.79+434835.7&53052&1736&139&7738&-5.47&0.04&DQ \\ 
        SDSS\,J084131.55+332915.6&52642&933&16&6810&-6.62&0.05&DQ\\ 
        SDSS\,J085239.66+042804.5&52670&1190&177&9555&&&DQ \\ 
        SDSS\,J085506.62+063904.7&52668&1189&535&7337&-5.99&0.01&DQ \\ 
        SDSS\,J085709.01+060357.4&52668&1189&27&8210&-5.00&0.02&DQ \\ 
        SDSS\,J090449.73+395416.4&52703&1199&595&7319&-5.74&0.05&DQ \\ 
        SDSS\,J090514.78+090426.5&52973&1300&420&8861&-4.89&0.01&DQ
        	(H$\alpha$ ?)\\ 
        SDSS\,J090632.17+470235.8&52606&898&565&5070&-4.31&0.01&DQ
	         (weak bands)\\ 
        SDSS\,J091602.73+101110.5&53050&1739&592&8715&-4.81&0.01&DQ\\ 
        SDSS\,J091830.27+484323.0&52637&900&429&8884&-3.72&0.06&DQ\\ 
        SDSS\,J092153.46+342136.9&52995&1274&216&8202&-5.46&0.09&DQ \\ 
        SDSS\,J092613.46+472521.1&52637&900&41&7261&-6.41&0.06&DQ \\ 
        SDSS\,J092909.03+331011.7&52991&1593&94&6361&-5.57&0.01&DQ\\ 
        SDSS\,J094014.65+090641.8&52993&1304&45&6169&-7.34&0.05&DQpec
            	(rounded bands)\\ 
        SDSS\,J094115.18+090154.4&52993&1304&9&9122&-4.73&0.02&DQ\\ 
        SDSS\,J094138.08+441458.2&52672&1202&58&8113&-5.42&0.06&DQ \\ 
        SDSS\,J095934.95+453725.4&52703&942&446&7211&-5.52&0.06&DQ \\ 
        SDSS\,J100059.82+100531.7&53053&1308&307&7958&-4.83&0.02&DQ, WD\,1105+412 \\ 
        SDSS\,J101750.38+373637.5&52996&1427&216&7497&-5.79&0.03&DQ \\ 
        SDSS\,J101800.00+083820.3&52762&1237&621&7784&-5.85&0.01&DQ\\ 
        SDSS\,J102635.81+580714.8&52316&559&6&8879&-4.56&0.03&DQ \\ 
        SDSS\,J110759.46+405910.9&53046&1437&512&7169&-6.45&0.02&DQ \\ 
        SDSS\,J110912.21+424956.0&53053&1363&37&9402&-4.84&0.08&DQ, WD\,0913+103 \\ 
        SDSS\,J112604.29+441938.6&53062&1365&564&7097&-6.38&0.05&DQ \\ 
        SDSS\,J113534.61+572451.7&53033&1310&485&7385&-6.42&0.02&
        	GD 311, \cite{Kawka.Vennes06} \\ 
        SDSS\,J115149.92+452729.8&53084&1368&503&8829&-4.75&0.04&DQ\\ 
        SDSS\,J122545.87+470613.0&53117&1451&35&6109&-5.82&0.01&DQ \\ 
        SDSS\,J123347.60+125346.1&53169&1616&429&7152&-6.36&0.02&DQ \\ 
        SDSS\,J130945.62+444541.0&53084&1375&391&8085&-4.01&0.03&DQ\\ 
        SDSS\,J131534.72+471108.9&53062&1461&428&7524&-5.99&0.01&DQ \\ 
        SDSS\,J131930.66+140137.1&53112&1773&105&7626&-5.65&0.05&DQ \\ 
        SDSS\,J133127.04+670419.5&51988&496&583&8899&-4.91&0.06&DQ
	        (weak bands)\\ 
        SDSS\,J143144.83+375011.9&53089&1381&599&6173&-6.97&0.02&DQ\\ 
        SDSS\,J152812.05+513445.2&52378&795&319&7531&-5.69&0.03&DQ \\ 
        SDSS\,J153447.54+414559.4&53149&1679&616&7804&-5.92&0.02&DQZ (CaII) \\ 
        SDSS\,J161653.36+392444.4&52759&1336&572&7319&-5.92&0.02&DQ \\ 
        SDSS\,J165436.86+315754.4&52791&1176&238&7258&-5.89&0.01&DQ \\ 
        SDSS\,J171341.76+324009.1&52413&976&623&7901&-5.36&0.01&DQ \\ 
        SDSS\,J211130.04$-$003628.8&52431&985&35&7168&-6.27&0.05&DQ\\ 
        SDSS\,J213503.32+000318.4&52468&989&198&6413&-6.78&0.03&DQ\\ 
	\noalign{\smallskip}
        \hline
      \end{tabular}
      \label{tabledq}
    \end{center}
\end{table*}

\begin{table*}[ht!]
  \caption{25 rediscovered DQ white dwarfs in the same format as
    Table~\ref{tabledq}. 20 of these are analyzed also in
    DB05. Information on the remaining 5 can be found in the
    \cite{McCook.Sion99} catalog (WD\,1426+613),
    \cite{Liebert.Harris.ea03}, \cite{Carollo.Bucciarelli.ea06}, and
    \cite{Harris.Liebert.ea03}. }
    \begin{center} \scriptsize
      \begin{tabular}{l|rrr|ccc|c}
        \hline
	\noalign{\smallskip}
        SDSS Name&MJD&Plate&Fid&$ \Teff [K]$& [C/He] &$ \sigma$
        [C/He]&Spectral Type/Ref\\ 
	\noalign{\smallskip}
        \hline
        SDSS\,J000011.57$-$085008.4&52143&650&450&8042&-5.46&0.08&\\ 
        SDSS\,J000807.54$-$103405.6&52141&651&199&7768&-5.66&0.06&\\ 
        SDSS\,J002531.50$-$110800.9&52145&653&86&8367&-4.96&0.01&\\ 
        SDSS\,J015433.57$-$004047.2&51871&403&268&7435&-5.89&0.03&\\ 
        SDSS\,J015441.75+140308.0&51877&430&558&6511&-6.89&0.01&\\ 
        SDSS\,J032054.11$-$071625.4&51924&460&236&6266&-5.45&\\ 
        SDSS\,J033218.22$-$003722.1&51810&415&240&8600&-4.62&0.03&\\ 
        SDSS\,J090157.92+575135.9&51924&483&600&&&&
                 	\cite{Liebert.Harris.ea03} \\ 
        SDSS\,J091922.18+023605.0&51929&473&458&11566&&&\\ 
        SDSS\,J093537.00+002422.0&52314&476&461&4958&-6.19&0.02&
                	\cite{Harris.Liebert.ea03} \\ 
        SDSS\,J094004.64+021022.6&52026&477&493&7283&-5.95&0.01&\\ 
        SDSS\,J095137.60+624348.7&51943&487&227&8388&-5.11&0.10&\\ 
        SDSS\,J113359.94+633113.2&52059&597&139&12082&&&\\ 
        SDSS\,J114851.68$-$012612.8&52056&329&578&9174&-3.73&0.01&\\ 
        SDSS\,J125359.61+013925.6&52026&523&252&8282&-4.98&0.02&\\ 
        SDSS\,J123752.12+415625.8&53090&1454&146&5846&-5.51&0.01&
	                \cite{Carollo.Bucciarelli.ea06} \\ 
        SDSS\,J132858.20+590851.0&52411&959&504&&&&
	                \cite{Liebert.Harris.ea03} \\ 
        SDSS\,J142728.30+611026.4&52368&607&379&6427&-6.83&0.01&WD\,1426+613  \\ 
        SDSS\,J144407.25+043446.8&52026&587&418&9449&-3.65&0.05&\\ 
        SDSS\,J144808.07$-$004755.9&51662&308&145&7063&-6.50&0.04&\\ 
        SDSS\,J154810.66+562647.7&52072&617&551&8119&-5.46&0.07&\\ 
        SDSS\,J155413.53+033634.5&52023&595&373&6512&-6.94&0.03&\\ 
        SDSS\,J164328.54+400204.3&52050&630&386&7144&-6.20&0.07&\\ 
        SDSS\,J165538.51+372247.1&52071&632&92&8997&-4.75&0.06&\\ 
        SDSS\,J205316.34$-$070204.3&52176&636&267&6382&-5.45&0.02&\\ 
	\noalign{\smallskip}
        \hline
      \end{tabular}
      \label{tabledq1}
    \end{center}
\end{table*}

Using this information about the position of DQs in the SDSS color
space, we have as a first selection step extracted from the SDSS
database all objects in the ``point source'' and ``unknown''
categories falling into the region defined by
\[  -0.8 < u-g < 1.5  \mbox{\qquad and \qquad}  -0.8 < g-r < 1.5,
 \]
resulting in 86856 selected objects. This sample still contains large
numbers of DA white dwarfs and quasars. We therefore have in a second
step refined the color region using a polygon in the $u-g, g-r$
plane. The polygon area follows rather closely the outline of the
theoretical grid, except at the hot end. Since the grid converges
towards a single line (the models are much less dependent on the C
abundance) the polygon at the hot end is significantly wider in
$u-g$. Applying this selection step narrowed the candidate sample down
to 44928 objects.

\subsection{Selection in 4-dimensional color space}
Since we consider for the moment only the dependence of colors on
\Teff\ and [C/He], it is obvious that the theoretical grid must define
a 2-dimensional surface in the 4-dimensional color space
($u-g,g-r,r-i,i-z$). That the variation is strongest in the first two
components suggest that this surface might be approximated by a plane
and this is confirmed by looking at different projections of the grid in
two- or three-dimensional subspaces. We have therefore fitted a plane
to the nodes of this grid using a $\chi^2$ minimization technique.
The fit is extremely good, with typical deviations of grid points from
the plane of 0.03 mag.

The equation of the plane is
\[
\left(\begin{array}{c} u-g \\g-r\\r-i\\i-z\end{array}\right)_{plane} =
\left( \begin{array}{r} 0.00003\\-0.00001\\-0.09854\\-0.18105
\end{array}\right)
 + (g-r) \left( \begin{array}{r}
  -0.00007\\1.00004\\0.53380\\0.19325\end{array}
\right) 
\]
\[  + (u-g) \left( \begin{array}{r}
  1.00000\\-0.00006\\-0.02747\\0.14871 \end{array} \right)
\] 
where we have used the first two colors directly as the two variables
for the plane. The reasonable region for DQs is given by the limits
\[
-0.55 < u-g < 1.3  \mbox{\qquad and \qquad} -0.3 < g-r < 1.4. \]

This plane representation of theoretical DQ colors was used in the
final photometric selection step. We defined a minimum ``distance''
$d$ from the plane for each observed object by measuring each
coordinate in units of their individual measurement errors
\[ 
     d^2      =
     \sum_{i=1}^{4}\left(\frac{c_{ob}(i)-c_{th}(i)}{\sigma(i)}\right)^2 .
\]
Here $c_{ob}$ and $c_{th}$ are the four observed and theoretical SDSS
colors, and $\sigma$ the measurement error of the observation.  We
eliminated the majority of objects by using a limit of $d=6.5$. This
limit had been determined from a test using the 40 DQs of
Fig.~\ref{figugr}. The final sample contained 6952 DQ candidates.

\subsection{Color space for different surface gravities}
We have also fitted the theoretical grids for \logg\ = 7.5 and 8.5 in
the same way. The resulting planes were practically indistinguishable
from the \logg\ = 8.0 plane, however, the lines of constant
temperature or C abundance are slightly shifted. This corresponds to
the well known fact for DQs that the parameters \logg\ and [C/H] are
degenerate. They cannot both be determined from photometry (nor
spectra), because a change of \logg\ can always be very nearly
compensated with a corresponding change of [C/He], and one of the two
has to be {\em assumed} to be able to proceed. Using the fact that
most white dwarfs cluster around 0.6~\Msun, we have assumed \logg\ =
8, as did DB05.

\subsection{Visual classification of the 6952 candidates}
Spectra for all remaining candidates were extracted from the SDSS
database and inspected visually. More than 95\%\ turned out to be
quasars, easily recognized by their strong and broad emission
lines. Of the remaining objects some were DA white dwarfs (broad
Balmer absorption), some galaxies with narrow absorption lines. The
DQs, the topic of this study, were identified through their molecular
Swan bands. In total we found 65 DQs, of which 25 were already known
in the literature and 37 are new detections, and 3 reclassifications
from a spectral type of DC.

Tables~\ref{tabledq} and \ref{tabledq1} list all 65 DQ with their SDSS
names, internal identifiers, and atmospheric parameters, determined as
described in the next section. Most of the objects show only carbon
features.  SDSS\,J090514.78+090426.5 shows a feature near the position
of H$\alpha$, but nothing near the other Balmer lines, and the
presence of hydrogen is highly uncertain. SDSS\,J153447.54+414559.4
shows Ca\,II H and K lines with equivalent widths of 2.1 and
1.0~\AA. The wings are broad and the distance of the star is very
likely less than 100~pc, implying a photospheric origin of the lines,
probably due to accretion. SDSS\,J094014.65+090641.8 shows rounded
bands similar to SDSS\,J223224.0-074434.3 in
\cite{Harris.Liebert.ea03}.

\begin{figure}
\includegraphics[width=8.8cm]{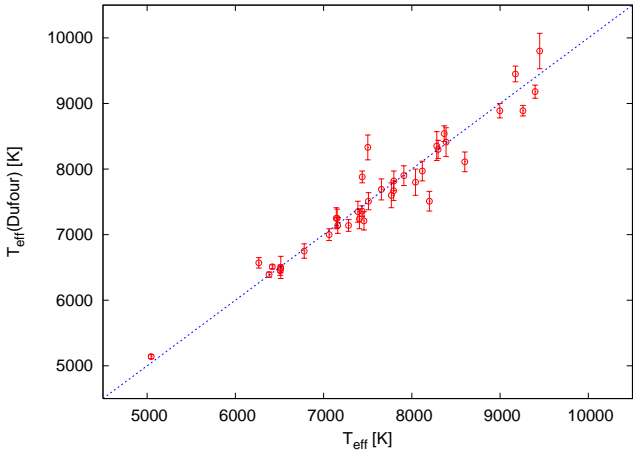}
\caption{Comparison of effective temperatures determined from the
  photometry with results from DB05
  \label{compt}}
\end{figure}

\section{Atmospheric analysis of the DQ white dwarfs}
\subsection{Photometry} 

The parameters \Teff\ and [C/He] were determined by minimizing in the
sense of a $\chi^2$ the distance between the observed point in
4-dimensional color space with model colors interpolated on the \Teff\
- [C/He] grid. $\chi^2$ minimization was obtained with the AMOEBA
routine from \cite{Press.Teukolsky.ea92}. Since our sample contains 20
objects in common with DB05, we compare the
derived \Teff\ in Fig.~\ref{compt}. The agreement in \Teff\ with those
from DB05, which are based on photometry and
spectra is excellent, confirming that photometry alone already gives a
reliable \Teff. From the differences between the two completely
independent determinations we estimate a typical error for \Teff\
of 190~K. This is a much more realistic error than the internal errors
from the $\chi^2$ routine; it is also very similar to the \Teff\ error
of 170~K cited by DB05, and we take this as our error estimate for
\Teff. On the other hand, the derived carbon
abundances differ markedly, with a much higher scatter and also
systematic differences. This is not really surprising, since the
colors for different abundances converge at the high temperature end
and a [C/He] determination based on colors alone will be inaccurate to
impossible. In addition, small uncertainties of the magnitude
zero-points could shift the whole grid by a few hundreds of a
magnitude, resulting in systematic changes. These abundances are not
used further and therefore not shown here.

\subsection{Spectroscopy}
Fortunately, SDSS provides photometry {\em and} spectroscopy. The
spectra were fitted with the theoretical spectra from our standard
grid using the same methods used by us for other types of white dwarfs
\citep{Koester.Napiwotzki.ea01, Homeier.Koester.ea98}. In principle
the spectroscopic fit provides \Teff\ and [C/He] (\logg\ = 8 is always
assumed). However, since we have shown the high reliability of the
photometric determination, we have kept the temperature fixed and
determined only the carbon abundance from the spectra, leading to
smaller errors of these abundances. These abundances are given in
Table~\ref{tabledq} and \ref{tabledq1}. For the five objects with
missing entries in the [C/He] column we could not obtain reasonable
fits within our grid. The spectral fits for the region around the
strongest Swan bands of the remaining 60 objects are shown in
Figure~\ref{specfit1} to Fig.~\ref{specfit3} in the appendix; the
values are also in Table~\ref{tabledq} and \ref{tabledq1}.

\begin{figure}
\includegraphics[width=8.8cm]{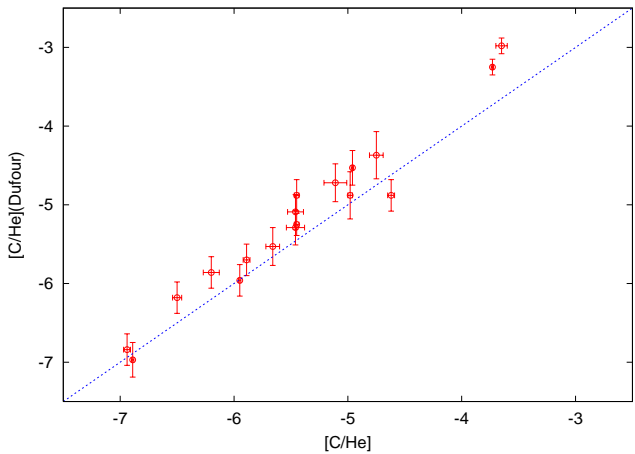}
\caption{comparison of carbon abundances from spectral fitting with
  results from DB05
  \label{compc}}
\end{figure}

Figure~\ref{compc} gives a comparison of our final results for the
carbon abundance with DB05 for the common
objects. The general agreement is very good, with a small systematic
shift of about 0.2 dex, increasing from lower to high abundances. The
[C/He] abundances of DB05 are on average slightly
higher than our values.

\section{Results and discussion}
We have identified 40 new DQ white dwarfs and rediscovered another 25
from the Data Release 4 of the Sloan Digital Sky Survey and presented
effective temperatures and carbon abundances for 60 of them. This
increases very significantly the number of DQs with a careful analysis
using up-to-date model atmospheres. Particularly important is the fact
that our sample is based on a completely homogenous set of photometry
and spectra. This is similar to the work of DB05, who however included
also observations from other sources together with 40 SDSS objects in
their sample.  For the objects in common with their analysis we find
good agreement for the parameters, with a slight offset in the
abundances of the order of 0.2 dex.
\begin{figure}
\includegraphics[width=8.8cm]{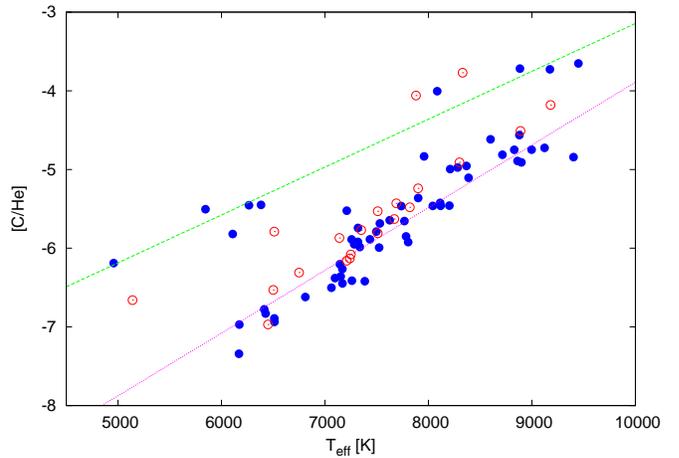}
\caption{Carbon abundance versus effective temperatures. Full circles are
  the results from this paper. We have added as open circles the results
  from DB05 for those objects not in our sample
  \label{comptc}}
\end{figure}

The major result of this paper is shown in Fig.~\ref{comptc}, which
displays the carbon abundance versus effective temperature. In
addition to our own results (shown as full circles) we have here
included data from DB05 for those objects not included in our sample
(open circles). We clearly confirm the three main conclusions of DB05:
\begin{itemize}
\item The paucity of objects with \Teff\ below 6000~K. Our sample adds
  only one object clearly below the apparent cutoff, also carbon-rich
  as the single cool objects in DB05.
\item The clear sequence for the majority of the objects, from low
  \Teff, low [C/He] to high \Teff, high [C/He], indicated by the lower
  of the two regression line fits (these fits use only the results
  from this paper). The existence of this  sequence
  was first pointed out by DB05 and shows the large advantage of using
  a homogenous sample of good quality observations compared to
  e.g. the study of \cite{Weidemann.Koester95}. The
  extrapolation of the sequence to 10000~K predicts [C/He] = -4.0, in
  agreement with the theoretical calculations of
  \cite{Fontaine.Brassard05} for a He-layer thickness of $10^{-4}$ of
  the stellar mass.
\item The existence of a separate population of carbon-rich DQs, with
  an abundance about 1 dex higher than the primary sequence. This
  population seems to form a second sequence, which we have
  tentatively described with the upper regression line.
\end{itemize}

Based on one single object with a parallax and thus mass determination
of 1.05~\Msun\ in the upper sequence (G47-18 = WD\,0856+33), DB05
tentatively suggested that all objects of this group are massive white
dwarfs and might be the descendents of the hot DQ stars, which also
seem to be massive, although that assertion is also only based on two
objects \citep{Liebert.Harris.ea03,
 MacDonald.Hernanz.ea98, Weidemann05}. Stellar evolution calculations
predict smaller He envelope masses for more massive progenitors
\citep{Kawai.Saio.ea88} and might lead to the dredge-up occurring at
higher effective temperatures \citep{Thejll.Shipman.ea90}. This would
also be a natural explanation for the absence of the high-mass tail in
the DB mass distribution \citep{Beauchamp.Wesemael.ea96,
Liebert.Harris.ea03}.

While these conclusions seem rather speculative at present, it is
obvious that further study of the large number of DQs with excellent
observations coming from SDSS and other large scale survey will
provide important clues for the remaining open question of the origin
and evolution of white dwarf surface compositions and spectral
types. Of particular importance are parallax determinations for DQs on
the high abundance sequence.

\acknowledgement{ This study was partially supported by a grant from
  the Deutsche Forschungsgemeinschaft (KO731/21-1,-2), and would have
  been impossible without the SDSS.  Funding for the Sloan Digital Sky
  Survey (SDSS) has been provided by the Alfred P. Sloan Foundation,
  the Participating Institutions, the National Aeronautics and Space
  Administration, the National Science Foundation, the U.S. Department
  of Energy, the Japanese Monbukagakusho, and the Max Planck
  Society. The SDSS Web site is http://www.sdss.org/.  The SDSS is
  managed by the Astrophysical Research Consortium (ARC) for the
  Participating Institutions. The Participating Institutions are The
  University of Chicago, Fermilab, the Institute for Advanced Study,
  the Japan Participation Group, The Johns Hopkins University, the
  Korean Scientist Group, Los Alamos National Laboratory, the
  Max-Planck-Institute for Astronomy (MPIA), the Max-Planck-Institute
  for Astrophysics (MPA), New Mexico State University, University of
  Pittsburgh, University of Portsmouth, Princeton University, the
  United States Naval Observatory, and the University of Washington.
  This research has made use of the SIMBAD database, operated at CDS,
    Strasbourg, France}


\appendix
\section{Graphical display of spectral fits for 60 DQs}
\begin{figure*}
\includegraphics[width=17.2cm]{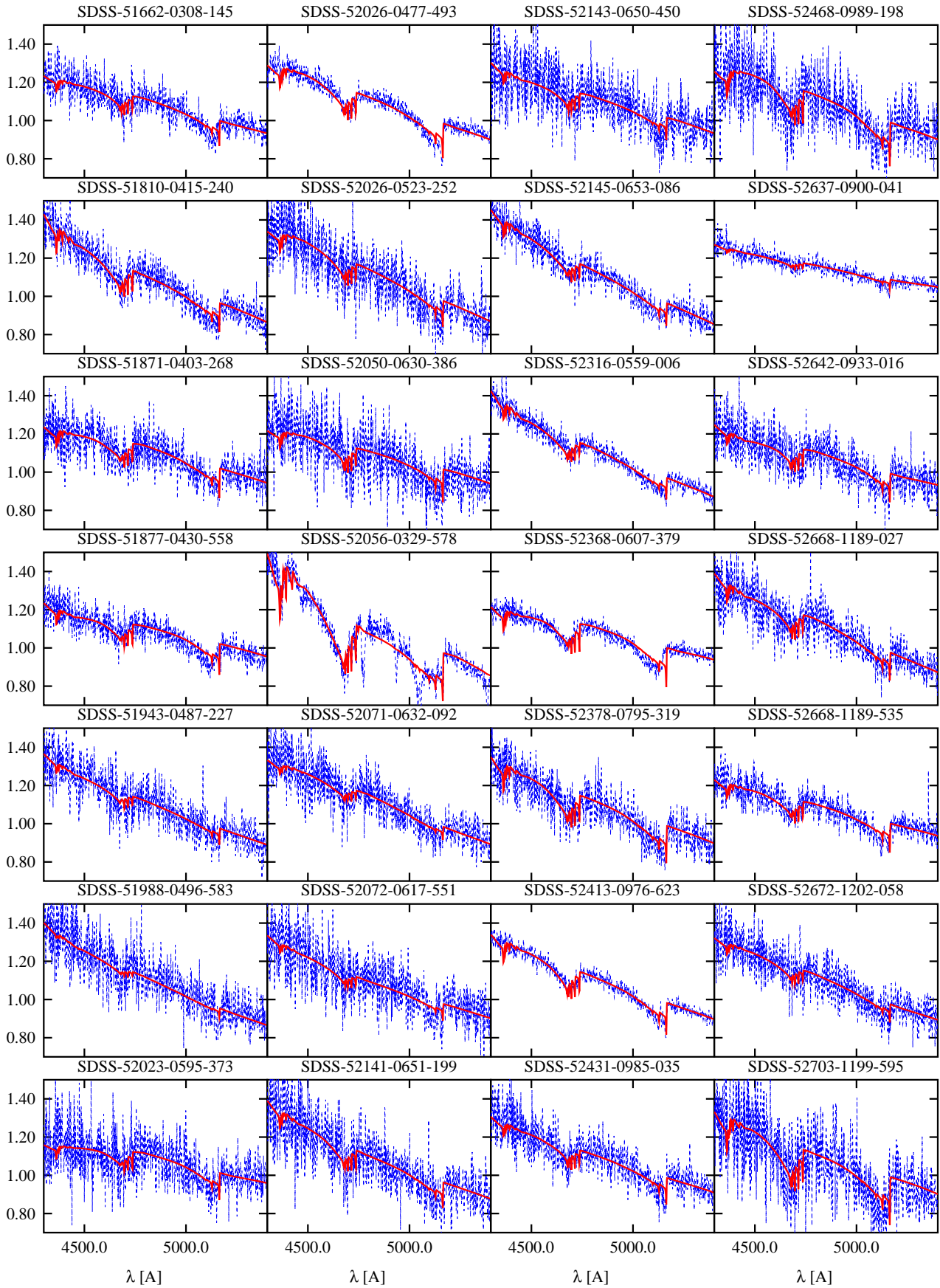}
\caption{Spectral fit for DQ white dwarfs. Thin (blue): observed;
  thick (red): model. Vertical axis is relative intensity on an
  arbitrary scale. \label{specfit1} }
\end{figure*}

\begin{figure*}
\includegraphics[width=17.5cm]{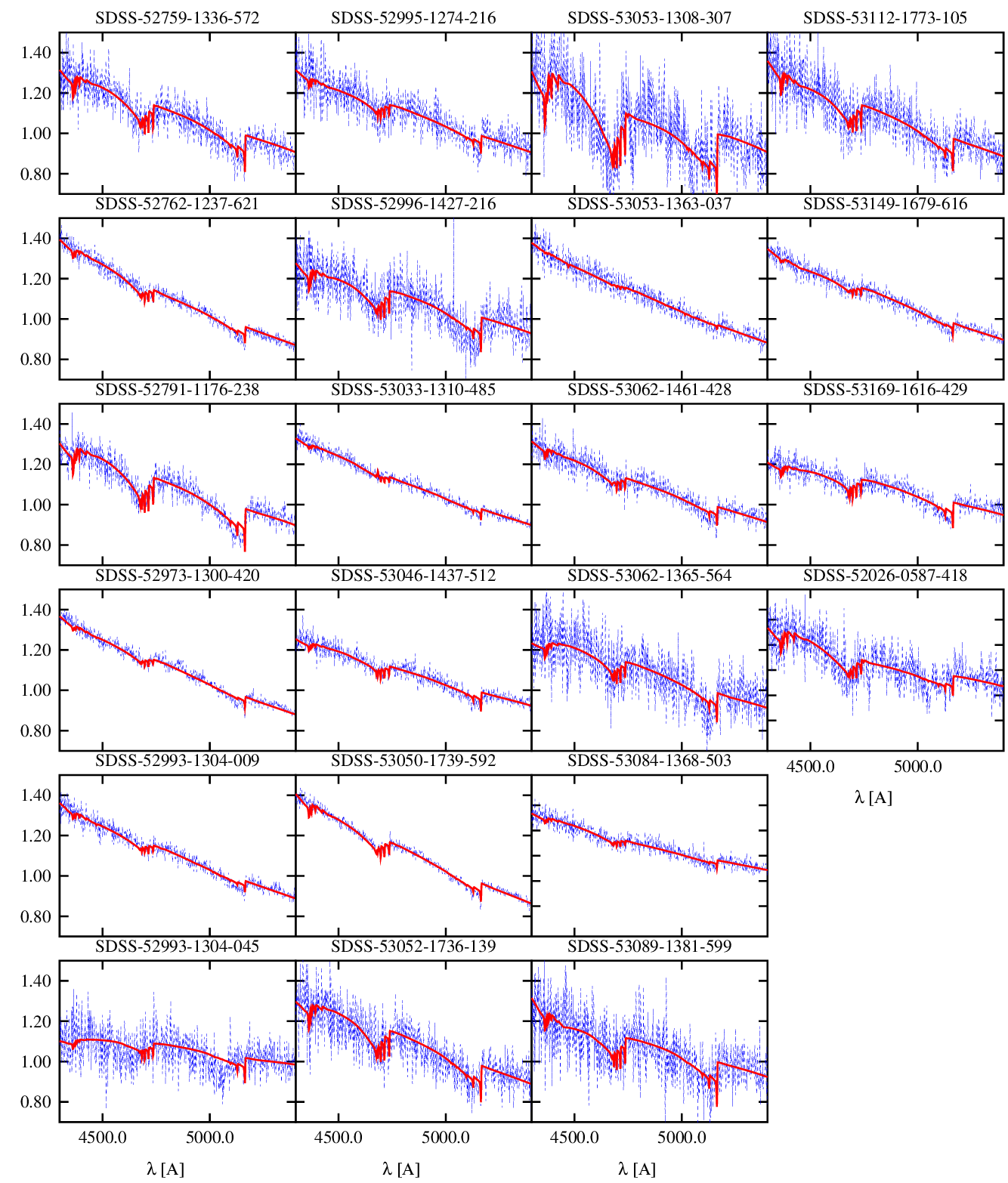}
\caption{Spectral fit for DQ white dwarfs; continued from
Fig.~\ref{specfit1} \label{specfit2} }
\end{figure*}

\begin{figure*}
\includegraphics[width=17.5cm]{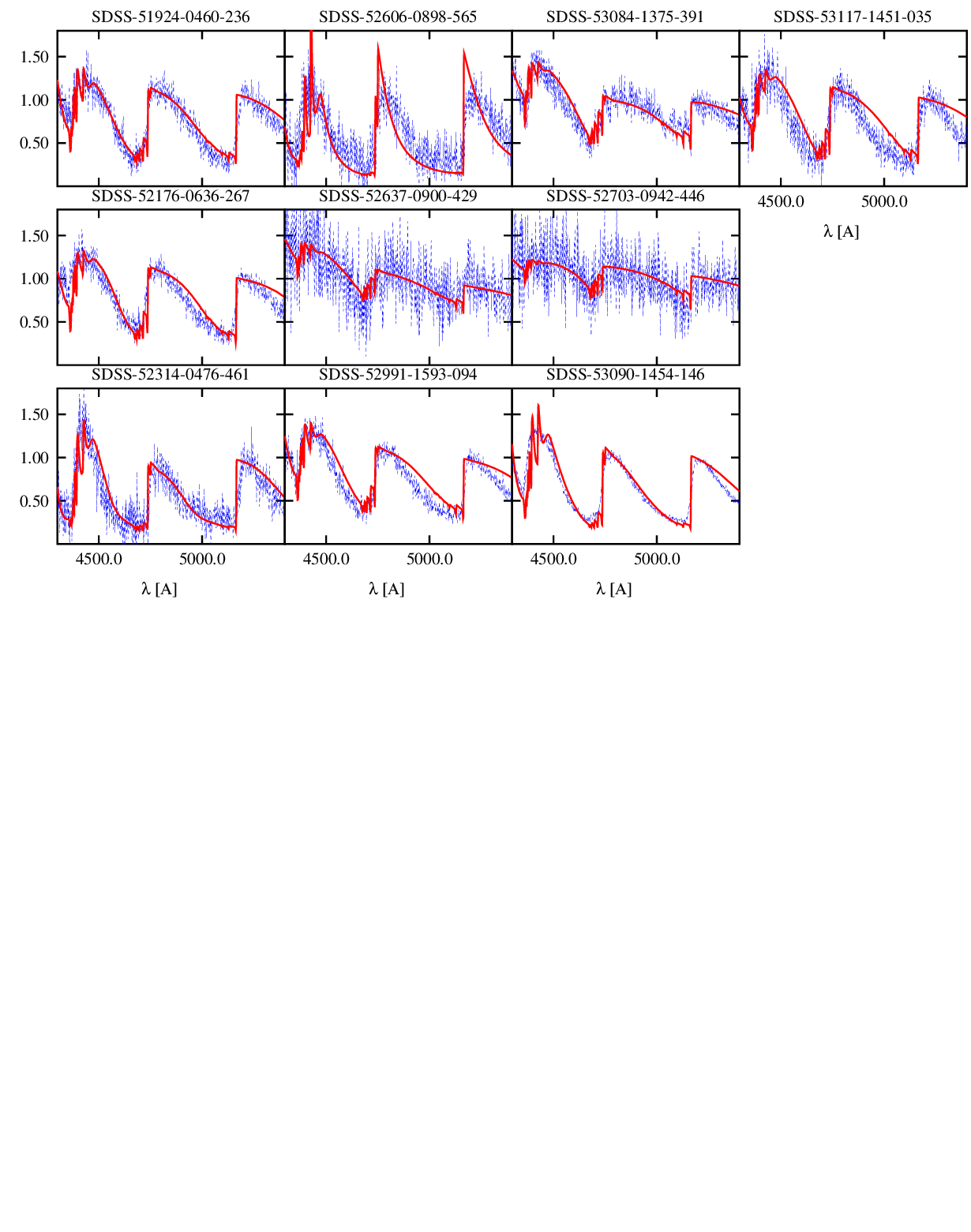}
\caption{Spectral fit for DQ white dwarfs; continued from
Fig.~\ref{specfit2} \label{specfit3} }
\end{figure*}

\end{document}